# Julian Schwinger:
# A Centennial Celebration at Harvard

by Steve Nadis

"Prodigies don't always pan out. Julian Schwinger did." Thus proclaimed an October 1996 article in Physics Today. "He had fastened onto physics for his life's work while still in his early teens when he got to the letter 'P' in a systematic odyssey through Encyclopedia Britannica." Schwinger was born in New York City on February 12, 1918, the son of Polish-Jewish émigrés whose families were in the garment industry, and on February 12, 2018, exactly 100 years after his birth (and 14 years after his death), his career and extraordinary accomplishments were celebrated at Harvard's Jefferson Laboratory.

There was indeed a lot to celebrate. Schwinger published a physics paper in a professional journal when he was just 16 years old—the first of more than 200 publications that followed. He enrolled in the City College of New York (CCNY) at the same age but came close to flunking out due to the large number of courses he was required to take outside of physics that were of little interest to him. Luckily, Schwinger crossed paths with Columbia physicist Isidor Rabi who immediately recognized the teenager's preternatural abilities. Schwinger then transferred to Columbia where he had the freedom to indulge his passion for physics. There, under Rabi's supervision, he earned a Bachelor's degree and PhD by the age of 21. Schwinger then worked for two years at Berkeley as J. Robert Oppenheimer's postdoctoral assistant before taking a lectureship at Purdue, interspersed with work at MIT's "Rad Lab" to help advance the nation's radar capabilities during World War II.

From 1945 to 1972, Schwinger served as a full-time member of the Harvard Physics faculty. In the course of an incredibly diverse career, he made important contributions to broad areas of physics including nuclear, atomic, particle and condensed matter physics, statistical mechanics, classical electromagnetism, synchrotron radiation, waveguide theory, general relativity, and quantum field theory. But he is best known, by far, for providing an

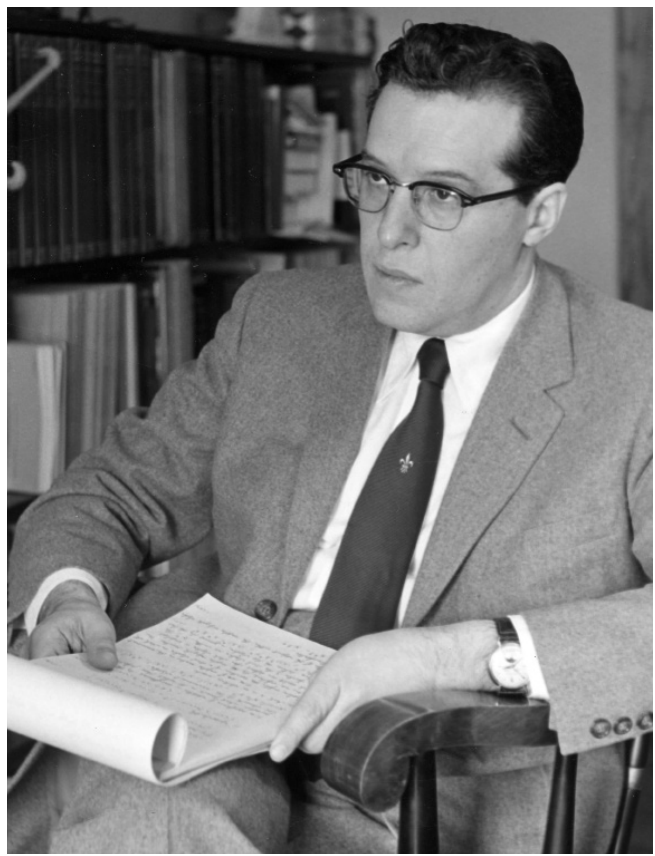

Image courtesy of AIP Emilio Segrè Visual Archives, Weber Collection.

essentially complete theory of quantum electrodynamics (QED), which combines quantum mechanics and special relativity to describe the interactions between light, matter, and the electromagnetic field. Schwinger unveiled his reformulation of QED in high profile lectures in 1948, as well as in a series of papers published in 1948 and 1949 in Physical Review, while continuing to expand upon these ideas in the 1950s. He shared the 1965 Nobel Prize for this work with Richard Feynman and Sin-Itiro Tomonaga, other key framers of QED theory. In 1951, Schwinger became (with Kurt Gödel) the first winner of the Albert Einstein Award; he earned the National Medal of Science in 1964.

It was this legacy that drew a standing-room-only crowd of more than two hundred attendees—including current physics students, faculty, and research scholars, plus former Schwinger students and other interested parties—to Jefferson Lab 250 in February of this year. People had come to pay tribute and learn about a man widely regarded as one of the 20th century's greatest physicists. Physics Professor Howard Georgi (BA '67) delivered the opening remarks for the proceedings called "Memories of Julian." He was joined onstage by three former Schwinger students—Sheldon Glashow (PhD '59), an emeritus professor of physics at Boston University and Harvard; Roy Glauber (BA '46, PhD '49), an emeritus professor of physics at Harvard; and Daniel Kleitman (PhD '58), an emeritus professor of applied mathematics at MIT—and a former assistant, Walter Gilbert, who previously served as a Harvard professor of molecular and cell biology. All told, Schwinger supervised 73 PhD recipients, making him one of the most prolific graduate advisors in physics anywhere. Like their mentor, Gilbert, Glashow, and Glauber each won a Nobel Prize, as did two other Schwinger students, Walter Kohn (PhD '48) and Benjamin Roy Mottleson (PhD '50).

Georgi kicked off the Special Colloquium by introducing himself as a "Schwinger grand-student," given that he did his PhD work at Yale under Schwinger student Charles Sommerfield (PhD '57). In his junior year as a Harvard undergraduate, Georgi took Schwinger's 253 (quantum field theory) course. "It was amazing," Georgi recalled. "He was a magisterial lecturer. He just had total control, not just of the material but of the total class… The blackboard was spectacular. Every mark was clearly planned, and he always ended at the right hand side of the blackboard in Jefferson 356 so that at the end of class he could scoot out. I always assumed he was escaping from his graduate students. But I'm looking forward to finding out what was actually going on with this mysterious character."

Georgi, fortunately, had some expert help with that inquiry, as his remarks were followed by those of the four highly credentialed panelists—Glauber, Gilbert, Kleitman, and Glashow—people who knew Schwinger well.

Some insights into Schwinger's character were also volunteered during the question-and-answer session by a former student, Fred Cooper (PhD '69), as he described the scene in Stockholm at the 1965 Nobel Prize award ceremony. "In a room full of people, including a large number of reporters," Cooper recounted, "Schwinger humbly told the audience: 'I woke up this morning, and the problems I couldn't solve yesterday, I can't solve today.'" Maybe the speedy exits from the lecture hall that Georgi alluded to were designed to avoid student questions so that Schwinger could attend, instead, to the questions bubbling within his own fertile mind, coupled with an urge to resume work on the intransigent problems he was still determined to solve.

Roy Glauber (excerpted remarks).

I thought I'd make a few observations about the history of the times we lived in. First, let me remind you that even the atomic nucleus is a relative newcomer on the scene. It dates from the turn of the last century, and very little was known about it for a long time. Many of the things that were found out about the nucleus were found out by one particular man, I. I. Rabi of Columbia. And he [Rabi] did one more thing: He found a shy CCNY "refugee" among the students at Columbia, and that was Julian Schwinger. Julian had a difficult time at CCNY; he went uptown to sit in on lectures at Columbia, and Rabi noticed him and acted as a kind of godfather. He, in fact, was responsible for giving Schwinger his PhD at Columbia and sending him to California to work with Oppenheimer. Schwinger spent most of the war years at MIT. I was at Los Alamos then, a bit underage I might add.

The time I want to focus on now is the end of the war when people were devoted to peacetime thoughts. How were we going to investigate the physics of these nuclei? The energies available were really quite low. Well at Los Alamos we had a lecture by a visitor in either September or October of 1945. It was a lecture on a new accelerator designed by Julian Schwinger, whoever he was. Bill Rarita happened to work in an office opposite mine and introduced me to this Schwinger chap who turned out to be rather short of stature and wore his hair with a bit of a pompadour up in front, which gained him about an inch in height. He was a rather shy guy, but he did give a lecture on a kind of particle accelerator he had designed, a proton accelerator. It was quite a clever and very simple device.

The extraordinary thing was the lecture. It lasted about an hour and a half and, I have to say, was one of the best lectures I ever heard. It was extraordinary because he had worked out every last detail of this device. I found it unbelievably impressive, especially compared to a great many of the lectures I heard at Los Alamos. No such lecture ever had the smoothness or the continuity or obvious cogency of this particular lecture.

At that stage, I still had two undergraduate courses in fields other than science that I had not yet taken, so I had no degree and felt I had little choice but to come back here [to Harvard]. Meanwhile, I had been figuring that I would go to work for my boss at Los Alamos, Hans Bethe, who was in his way enormously impressive and a real father figure. I have to say that hearing this single lecture from Schwinger and given the knowledge that he had just received an appointment at Harvard, I was seriously shaken in my determination to go back with Hans Bethe when he returned to Cornell, and in fact I didn't. I came back here, absolutely delighted that Schwinger was going to be at Harvard from that point on.

Walter Gilbert (excerpted remarks):

I had a postdoctoral fellowship at Harvard [starting in 1957], and the year after that Harvard made me Julian's assistant. Julian and I constantly talked about doing something together and then never did. Our entire experience was actually going to Julian's lectures.

[The students and I] sat entranced as he covered the blackboard from one top corner to the lower further corner and copied everything down, thinking that at least by writing it down we would be able to catch some of the mystery and some of the magic of it. In those years he was lecturing on his conception of quantum mechanics, which he called measurement theory, and it was wonderfully obscure. I never could do anything with it.

The physics world centered around Julian in a curious way. Almost all of us were doing other things in physics and talking to him off and on. I remember spending a long afternoon standing outside the building here, leaning on the cars with Julian, discussing the Hungarian uprising and the world situation. It was probably the only serious conversation I had with him throughout that entire period. And I carried it in my memory. After every lecture, we would go down to the restaurant in Harvard where we could have a 99 cent lunch. It had to be under a dollar for tax reasons.

Daniel Kleitman (excerpted remarks):

Sheldon Glashow and I graduated from Cornell University in 1954 and enrolled as graduate students in the physics department at Harvard. We both enrolled in Professor Schwinger's course, among others, and continued to take his courses for the next two years. He started from the very beginning with his formulation of quantum mechanics and proceeded to develop quantum field theory and much more.

The class consisted of about 30 students assembled at exactly five minutes after the hour. Schwinger would arrive at the door and immediately begin his lecture. He spoke without notes and talked in a quiet voice in a manner so crystal clear and persuasive that it was hypnotic. Nobody dared to ask a question. Aside from his voice and the sound of his chalk, there was absolute silence in the room aside from occasional burps from a classmate who had a digestive disorder. At exactly five minutes before the hour, he would be positioned close to the door, would put down his chalk and exit, often moving rapidly toward his sports car with which he would travel home. The lectures were as well organized and informative as any I have ever experienced. Every one of them.

As an adviser, he was always friendly and helpful. My problem was I thought that the time to talk with him was when I made significant progress. Much later I learned that the time for a student to talk to his adviser was when he was stuck. Fortunately, I was able to develop a working relation with this gentleman here, Professor Roy Glauber, and with his aid and comfort and the kindness of Julian Schwinger, I graduated and went out into the world.

Julian Schwinger's lectures were wonderful and inspiring. They inspired us to do something similar—to produce out of one's own mind a reformulation of an important subject that will solve important problems. I never found the opportunity to do that in the world of physics. Fortunately for me, I was lucky enough to do such things at a much, much smaller scale in mathematics, and so I became a mathematician.

Sheldon Glashow (excerpted remarks):

I was at Cornell [as an undergraduate] because I was rejected at Harvard. But Harvard accepted me [for graduate school]. I guess we were in our second year that [Schwinger] had that famous interview with about ten of us: Danny [Kleitman] and me, Marshall Baker, Charlie Sommerfield, Ray Sawyer, and three guys whose name begins with "W," and maybe somebody else. By the time Schwinger got to me, he had run out of problems that made any sense, so he developed a remark that he had published earlier. He said

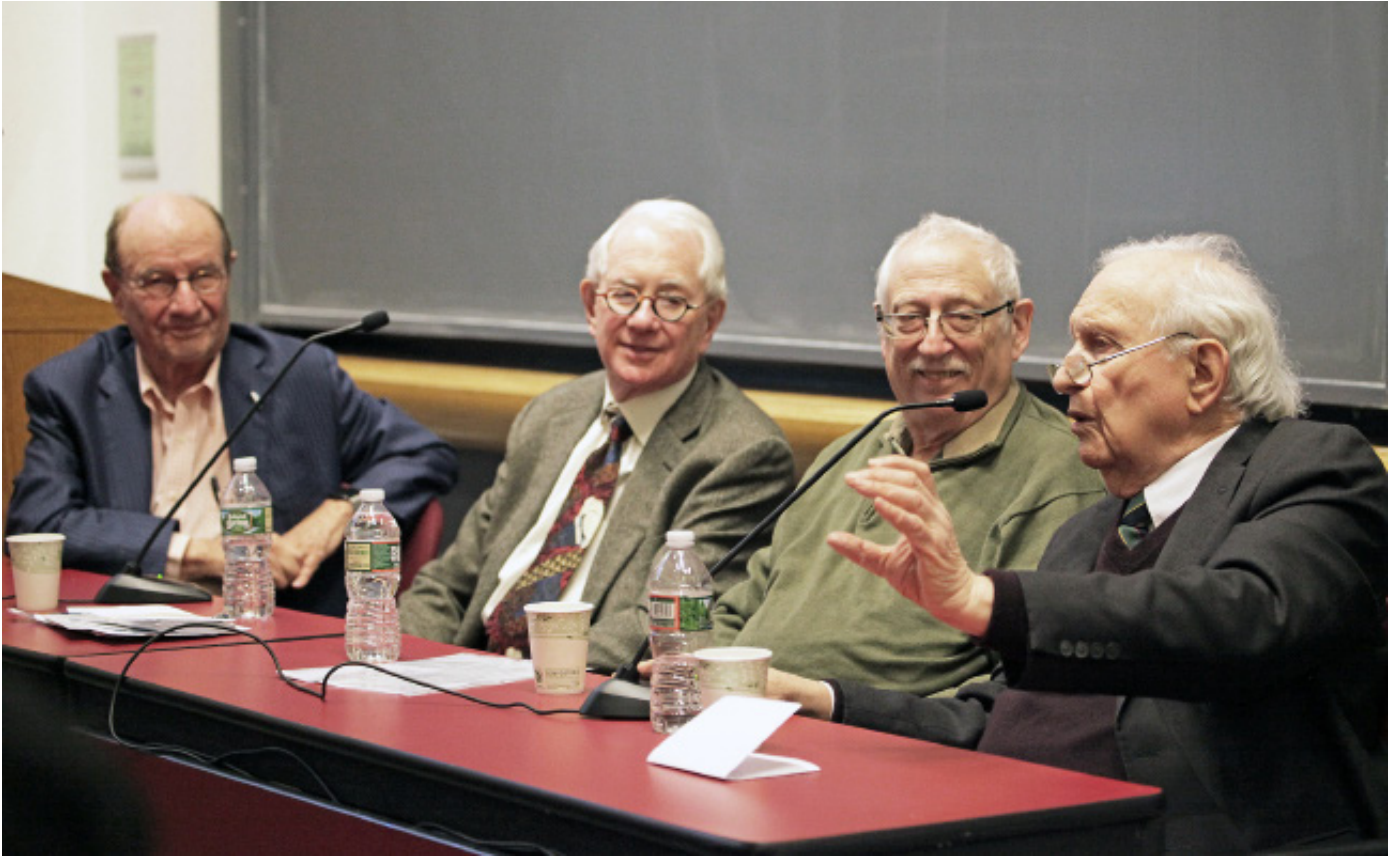

Left to right: Walter Gilbert, Sheldon Glashow, Daniel Kleitman, and Roy Glauber.  Photograph by Paul Horowitz.

that there are these things called Yang-Mills theories and they might be useful. "Why don't you think about that?" he asked. Anyway, I'd like to say right now that Schwinger was indeed the first person to invoke Yang-Mills theories, gauge theories, as we call them today, to unify weak and electromagnetic interactions. I found no such allegation anywhere in the literature aside from in his 1956 paper. So he sent me off and said do it. Of course I had no idea of how to do it.

Let me jump ahead. My PhD examination committee consisted of Paul Martin from here, recently sadly deceased, Julian Schwinger, a physicist named Sacks (a good friend of Julian's who was chairman of the department at Wisconsin at the time), and Frank Yang of Yang-Lee [and Yang-Mills fame]. I started to explain what I had done, which wasn't all that much, but I started by explaining how the electron neutrino and muon neutrino are very likely different from one another—different particles and that had to be built into the model. At that point Yang said, "Stop. Mr. Glashow, what do you mean the electron neutrino and muon neutrino are different from one another? There's no way to establish such a fact." And Julian said, "Shelly, quiet down. Let me answer Yang." So at that point my exam was more or less over, and Schwinger explained in great detail what such an experiment would be like. It would be the experiment that would be done later by Schwartz and Lederman and Steinberger.  He described how such an experiment could prove that electron neutrinos were different from muon neutrinos, and he, Schwinger, had very peculiar reasons, correct ultimately, that the neutrinos had to be different from one other another.

But I got my degree. And it was wonderful working with Julian. The one regret that we had, which we expressed to one another much later, was that we never got around to writing that paper on the electroweak theory that we should have written.